\documentclass[%
 amsmath,amssymb,
 aps,
]{revtex4-2}

\usepackage{graphicx}
\usepackage{dcolumn}
\usepackage{bm}
\usepackage[dvipsnames]{xcolor}
\usepackage[normalem]{ulem}
\usepackage[title]{appendix}
\usepackage{comment}
\usepackage{float}
\usepackage{mathtools}

\usepackage{hyperref}

\graphicspath{{./figures/}}
\begin{document}
		
			\title{\bf {Reply to comment on ``Failure of the simultaneous block diagonalization technique applied to complete and cluster synchronization of random networks''}}

			\author{ Shirin Panahi, Nelson Amaya, Isaac Klickstein, Galen Novello, Francesco Sorrentino \\
			University of New Mexico, Albuquerque, US 80131}

\begin{abstract}
   We respond briefly to a comment \cite{zhang2021comment} recently posted online on our paper \cite{panahi2022failure}. Complete and cluster synchronization of random networks is undoubtedly a topic of interest in the Physics, Engineering, and Nonlinear Dynamics literature.  In \cite{panahi2021cluster} we study both complete and cluster synchronization of networks and introduce indices that measure success (or failure) of application of the SBD technique in decoupling the stability problem into problems of lower dimensionality. Our usage of the word `failure' indicates  that the technique does not produce a decomposition which results in a system which is easier to analyze, not that the technique fails in correctly decoupling these problems. 
\end{abstract}

	\maketitle

This brief reply is in response to a comment \cite{zhang2021comment} recently posted online on our paper \cite{panahi2022failure}. Synchronization of networks of coupled oscillators has received considerable attention in the Physics, Engineering, and Nonlinear Dynamics literature \cite{Report,li2003synchronization,li2013control,xiang2016synchronization,scholl2016synchronization,boccaletti2018synchronization,totz2018spiral}. Following the original work by Pecora and Carroll \cite{Pe:Ca}, several approaches have been developed to reduce the dimensionality of the stability analysis for the synchronous solution of arbitrary networks. Among these, there are the techniques for simultaneous block diagonalization (SBD) of a set of matrices  \cite{UHLIG1973281,Maehara2010-1,Maehara2010-2,Murota2010,Maehara2011}, which were originally applied to problems in the areas of semidefinite programming and signal processing (independent component analysis), see e.g. \cite{Maehara2010-1}. 
The first application of these techniques to network synchronization was presented in a 2012 paper \cite{Ir:So}. Only recently they have been applied to the problem of cluster synchronization of networks \cite{Zhang2020,panahi2021cluster,zhang2021unified}. Our recent paper \cite{panahi2022failure} investigates application of SBD to stability of both complete and cluster synchronization in random (generic) 
networks. For both these problems, an index is defined that measures success (or failure) of application of the SBD technique in decoupling the stability problem into problems of lower dimensionality. In the case of random networks
 the extent of the dimensionality reduction achievable is very low and  equal to that produced by application of a trivial transformation.

In response to \cite{zhang2021comment}, we emphasize that it is useful to know the limitations of analytical tools. In \cite{panahi2022failure} we consider networks that display both complete and cluster synchronization, but for which application of the SBD techniques does not reduce the stability problem in a meaningful way. Therefore, it is important to recognize that from a practical standpoint analysis of these networks is still essentially an open problem.

In \cite{panahi2021cluster}, when we speak of the failure of the SBD technique applied to the study of various networks, what we refer to is the failure of the technique to produce a decomposition which results in a system which is easier to analyze.   In particular, we do not mean to imply that the methods fail from the standpoint of actually producing an SBD -- just that in practice this SBD may not actually be any easier to analyze than the original system. %
It seems that certain systems are essentially beyond the scope of the analysis possible with the tools currently available.  Again we would like to stress that we believe knowing the limitations of our tools is a good and useful thing.

That having being said, we think our paper states clearly why it is important to study random networks and what we exactly mean by `failure'. Below we excerpt two paragraphs from \cite{panahi2022failure}, which we think clearly explain the exact context of our words and the importance of our conclusions.\\

\textbf{On the importance of random networks}\\

{Random networks are broadly studied in the literature as fundamental and paradigmatic models for the structure and dynamics of complex systems \cite{Report}. Previous work has investigated random networks in the context of epidemics \cite{pastor2001epidemic,marder2007dynamics,pastor2015epidemic}, percolation \cite{achlioptas2009explosive,friedman2009construction}, resilience to attacks and failures \cite{guillaume2004comparison,liu2012cascading}, games \cite{devlin2009evolution}, network synchronization \cite{restrepo2006synchronization} and control \cite{liu2011controllability}.}
{It is therefore important to characterize both complete and cluster synchronization for this class of networks.} We show that application of the SBD reduction to these random networks does not lead to a beneficial reduction of the stability problem, either in the case of complete synchronization or cluster synchronization. Nonetheless, we do not mean that the technique is not useful, but that its usefulness is limited to the non-generic case, for which the reduction can sometimes be very significant\cite{Ir:So,Zhang2020}.\\


\textbf{On the extent of the dimensionality reduction}\\


In this paper we take the approach of the natural scientist and focus on whether or not a mathematical tool (the SBD decomposition) is effective in dealing with the synchronization of random networks. Ref. \cite{zhang2021comment} 
takes a different perspective and claims that random networks are not a good testbed for application of the SBD technique. Here we are interested in assessing whether problems of practical interest can be successfully addressed by the SBD tool, rather than  looking for problems to which the tool can be successfully, or rather conveniently, applied. 
Previous work in this area has often only emphasized the strengths and not the limitations of the technique, which is partially corrected in this paper. The fact that the technique mostly fails when applied to random networks points out the importance of developing alternative tools and/or new techniques to deal with the important class of random networks. A relevant related question is whether the SBD technique can be successfully applied to the analysis of real network topologies. This question has been recently considered in \cite{panahi2021cluster}
, which has shown a moderate success of the SBD technique in this case.

In closing we would like to thank Zhang \emph{et al} for their contributions to the field. 
We appreciate the feedback on our work and their role as contributors to this field in general. 

 \newcommand{\noop}[1]{}


\begin{thebibliography}{29}%
\makeatletter
\providecommand \@ifxundefined [1]{%
 \@ifx{#1\undefined}
}%
\providecommand \@ifnum [1]{%
 \ifnum #1\expandafter \@firstoftwo
 \else \expandafter \@secondoftwo
 \fi
}%
\providecommand \@ifx [1]{%
 \ifx #1\expandafter \@firstoftwo
 \else \expandafter \@secondoftwo
 \fi
}%
\providecommand \natexlab [1]{#1}%
\providecommand \enquote  [1]{``#1''}%
\providecommand \bibnamefont  [1]{#1}%
\providecommand \bibfnamefont [1]{#1}%
\providecommand \citenamefont [1]{#1}%
\providecommand \href@noop [0]{\@secondoftwo}%
\providecommand \href [0]{\begingroup \@sanitize@url \@href}%
\providecommand \@href[1]{\@@startlink{#1}\@@href}%
\providecommand \@@href[1]{\endgroup#1\@@endlink}%
\providecommand \@sanitize@url [0]{\catcode `\\12\catcode `\$12\catcode
  `\&12\catcode `\#12\catcode `\^12\catcode `\_12\catcode `\%12\relax}%
\providecommand \@@startlink[1]{}%
\providecommand \@@endlink[0]{}%
\providecommand \url  [0]{\begingroup\@sanitize@url \@url }%
\providecommand \@url [1]{\endgroup\@href {#1}{\urlprefix }}%
\providecommand \urlprefix  [0]{URL }%
\providecommand \Eprint [0]{\href }%
\providecommand \doibase [0]{https://doi.org/}%
\providecommand \selectlanguage [0]{\@gobble}%
\providecommand \bibinfo  [0]{\@secondoftwo}%
\providecommand \bibfield  [0]{\@secondoftwo}%
\providecommand \translation [1]{[#1]}%
\providecommand \BibitemOpen [0]{}%
\providecommand \bibitemStop [0]{}%
\providecommand \bibitemNoStop [0]{.\EOS\space}%
\providecommand \EOS [0]{\spacefactor3000\relax}%
\providecommand \BibitemShut  [1]{\csname bibitem#1\endcsname}%
\let\auto@bib@innerbib\@empty
\bibitem [{\citenamefont {Zhang}(2021)}]{zhang2021comment}%
  \BibitemOpen
  \bibfield  {author} {\bibinfo {author} {\bibfnamefont {Y.}~\bibnamefont
  {Zhang}},\ }\bibfield  {title} {\bibinfo {title} {Comment on" failure of the
  simultaneous block diagonalization technique applied to complete and cluster
  synchronization of random networks"},\ }\href@noop {} {\bibfield  {journal}
  {\bibinfo  {journal} {arXiv preprint arXiv:2110.15493}\ } (\bibinfo {year}
  {2021})}\BibitemShut {NoStop}%
\bibitem [{\citenamefont {Panahi}\ \emph {et~al.}(2022)\citenamefont {Panahi},
  \citenamefont {Amaya}, \citenamefont {Klickstein}, \citenamefont {Novello},\
  and\ \citenamefont {Sorrentino}}]{panahi2022failure}%
  \BibitemOpen
  \bibfield  {author} {\bibinfo {author} {\bibfnamefont {S.}~\bibnamefont
  {Panahi}}, \bibinfo {author} {\bibfnamefont {N.}~\bibnamefont {Amaya}},
  \bibinfo {author} {\bibfnamefont {I.}~\bibnamefont {Klickstein}}, \bibinfo
  {author} {\bibfnamefont {G.}~\bibnamefont {Novello}},\ and\ \bibinfo {author}
  {\bibfnamefont {F.}~\bibnamefont {Sorrentino}},\ }\bibfield  {title}
  {\bibinfo {title} {Failure of the simultaneous block diagonalization
  technique applied to complete and cluster synchronization of random
  networks},\ }\href@noop {} {\bibfield  {journal} {\bibinfo  {journal}
  {Physical Review E}\ }\textbf {\bibinfo {volume} {105}},\ \bibinfo {pages}
  {014313} (\bibinfo {year} {2022})}\BibitemShut {NoStop}%
\bibitem [{\citenamefont {Panahi}\ \emph {et~al.}(2021)\citenamefont {Panahi},
  \citenamefont {Klickstein},\ and\ \citenamefont
  {Sorrentino}}]{panahi2021cluster}%
  \BibitemOpen
  \bibfield  {author} {\bibinfo {author} {\bibfnamefont {S.}~\bibnamefont
  {Panahi}}, \bibinfo {author} {\bibfnamefont {I.}~\bibnamefont {Klickstein}},\
  and\ \bibinfo {author} {\bibfnamefont {F.}~\bibnamefont {Sorrentino}},\
  }\bibfield  {title} {\bibinfo {title} {Cluster synchronization of networks
  via a canonical transformation for simultaneous block diagonalization of
  matrices},\ }\href@noop {} {\bibfield  {journal} {\bibinfo  {journal} {Chaos:
  An Interdisciplinary Journal of Nonlinear Science}\ }\textbf {\bibinfo
  {volume} {31}},\ \bibinfo {pages} {111102} (\bibinfo {year}
  {2021})}\BibitemShut {NoStop}%
\bibitem [{\citenamefont {Boccaletti}\ \emph {et~al.}(2006)\citenamefont
  {Boccaletti}, \citenamefont {Latora}, \citenamefont {Moreno}, \citenamefont
  {Chavez},\ and\ \citenamefont {Hwang}}]{Report}%
  \BibitemOpen
  \bibfield  {author} {\bibinfo {author} {\bibfnamefont {S.}~\bibnamefont
  {Boccaletti}}, \bibinfo {author} {\bibfnamefont {V.}~\bibnamefont {Latora}},
  \bibinfo {author} {\bibfnamefont {Y.}~\bibnamefont {Moreno}}, \bibinfo
  {author} {\bibfnamefont {M.}~\bibnamefont {Chavez}},\ and\ \bibinfo {author}
  {\bibfnamefont {D.~U.}\ \bibnamefont {Hwang}},\ }\bibfield  {title} {\bibinfo
  {title} {Complex networks: Structure and dynamics},\ }\href@noop {}
  {\bibfield  {journal} {\bibinfo  {journal} {Phys. Rep.}\ }\textbf {\bibinfo
  {volume} {424}},\ \bibinfo {pages} {175} (\bibinfo {year}
  {2006})}\BibitemShut {NoStop}%
\bibitem [{\citenamefont {Li}\ and\ \citenamefont
  {Chen}(2003)}]{li2003synchronization}%
  \BibitemOpen
  \bibfield  {author} {\bibinfo {author} {\bibfnamefont {X.}~\bibnamefont
  {Li}}\ and\ \bibinfo {author} {\bibfnamefont {G.}~\bibnamefont {Chen}},\
  }\bibfield  {title} {\bibinfo {title} {Synchronization and desynchronization
  of complex dynamical networks: an engineering viewpoint},\ }\href@noop {}
  {\bibfield  {journal} {\bibinfo  {journal} {IEEE Transactions on Circuits and
  Systems I: Fundamental Theory and Applications}\ }\textbf {\bibinfo {volume}
  {50}},\ \bibinfo {pages} {1381} (\bibinfo {year} {2003})}\BibitemShut
  {NoStop}%
\bibitem [{\citenamefont {Li}\ \emph {et~al.}(2013)\citenamefont {Li},
  \citenamefont {Dasanayake},\ and\ \citenamefont {Ruths}}]{li2013control}%
  \BibitemOpen
  \bibfield  {author} {\bibinfo {author} {\bibfnamefont {J.-S.}\ \bibnamefont
  {Li}}, \bibinfo {author} {\bibfnamefont {I.}~\bibnamefont {Dasanayake}},\
  and\ \bibinfo {author} {\bibfnamefont {J.}~\bibnamefont {Ruths}},\ }\bibfield
   {title} {\bibinfo {title} {Control and synchronization of neuron
  ensembles},\ }\href@noop {} {\bibfield  {journal} {\bibinfo  {journal} {IEEE
  Transactions on automatic control}\ }\textbf {\bibinfo {volume} {58}},\
  \bibinfo {pages} {1919} (\bibinfo {year} {2013})}\BibitemShut {NoStop}%
\bibitem [{\citenamefont {Xiang}\ \emph {et~al.}(2016)\citenamefont {Xiang},
  \citenamefont {Wen},\ and\ \citenamefont {Pan}}]{xiang2016synchronization}%
  \BibitemOpen
  \bibfield  {author} {\bibinfo {author} {\bibfnamefont {S.}~\bibnamefont
  {Xiang}}, \bibinfo {author} {\bibfnamefont {A.}~\bibnamefont {Wen}},\ and\
  \bibinfo {author} {\bibfnamefont {W.}~\bibnamefont {Pan}},\ }\bibfield
  {title} {\bibinfo {title} {Synchronization regime of star-type laser network
  with heterogeneous coupling delays},\ }\href@noop {} {\bibfield  {journal}
  {\bibinfo  {journal} {IEEE Photonics Technology Letters}\ }\textbf {\bibinfo
  {volume} {28}},\ \bibinfo {pages} {1988} (\bibinfo {year}
  {2016})}\BibitemShut {NoStop}%
\bibitem [{\citenamefont {Sch{\"o}ll}(2016)}]{scholl2016synchronization}%
  \BibitemOpen
  \bibfield  {author} {\bibinfo {author} {\bibfnamefont {E.}~\bibnamefont
  {Sch{\"o}ll}},\ }\bibfield  {title} {\bibinfo {title} {Synchronization
  patterns and chimera states in complex networks: Interplay of topology and
  dynamics},\ }\href@noop {} {\bibfield  {journal} {\bibinfo  {journal} {The
  European Physical Journal Special Topics}\ }\textbf {\bibinfo {volume}
  {225}},\ \bibinfo {pages} {891} (\bibinfo {year} {2016})}\BibitemShut
  {NoStop}%
\bibitem [{\citenamefont {Boccaletti}\ \emph {et~al.}(2018)\citenamefont
  {Boccaletti}, \citenamefont {Pisarchik}, \citenamefont {Del~Genio},\ and\
  \citenamefont {Amann}}]{boccaletti2018synchronization}%
  \BibitemOpen
  \bibfield  {author} {\bibinfo {author} {\bibfnamefont {S.}~\bibnamefont
  {Boccaletti}}, \bibinfo {author} {\bibfnamefont {A.~N.}\ \bibnamefont
  {Pisarchik}}, \bibinfo {author} {\bibfnamefont {C.~I.}\ \bibnamefont
  {Del~Genio}},\ and\ \bibinfo {author} {\bibfnamefont {A.}~\bibnamefont
  {Amann}},\ }\href@noop {} {\emph {\bibinfo {title} {Synchronization: from
  coupled systems to complex networks}}}\ (\bibinfo  {publisher} {Cambridge
  University Press},\ \bibinfo {year} {2018})\BibitemShut {NoStop}%
\bibitem [{\citenamefont {Totz}\ \emph {et~al.}(2018)\citenamefont {Totz},
  \citenamefont {Rode}, \citenamefont {Tinsley}, \citenamefont {Showalter},\
  and\ \citenamefont {Engel}}]{totz2018spiral}%
  \BibitemOpen
  \bibfield  {author} {\bibinfo {author} {\bibfnamefont {J.~F.}\ \bibnamefont
  {Totz}}, \bibinfo {author} {\bibfnamefont {J.}~\bibnamefont {Rode}}, \bibinfo
  {author} {\bibfnamefont {M.~R.}\ \bibnamefont {Tinsley}}, \bibinfo {author}
  {\bibfnamefont {K.}~\bibnamefont {Showalter}},\ and\ \bibinfo {author}
  {\bibfnamefont {H.}~\bibnamefont {Engel}},\ }\bibfield  {title} {\bibinfo
  {title} {Spiral wave chimera states in large populations of coupled chemical
  oscillators},\ }\href@noop {} {\bibfield  {journal} {\bibinfo  {journal}
  {Nature Physics}\ }\textbf {\bibinfo {volume} {14}},\ \bibinfo {pages} {282}
  (\bibinfo {year} {2018})}\BibitemShut {NoStop}%
\bibitem [{\citenamefont {Pecora}\ and\ \citenamefont {Carroll}(1998)}]{Pe:Ca}%
  \BibitemOpen
  \bibfield  {author} {\bibinfo {author} {\bibfnamefont {L.}~\bibnamefont
  {Pecora}}\ and\ \bibinfo {author} {\bibfnamefont {T.}~\bibnamefont
  {Carroll}},\ }\bibfield  {title} {\bibinfo {title} {Master stability
  functions for synchronized coupled systems},\ }\href@noop {} {\bibfield
  {journal} {\bibinfo  {journal} {Phys. Rev. Lett.}\ }\textbf {\bibinfo
  {volume} {80}},\ \bibinfo {pages} {2109} (\bibinfo {year}
  {1998})}\BibitemShut {NoStop}%
\bibitem [{\citenamefont {Uhlig}(1973)}]{UHLIG1973281}%
  \BibitemOpen
  \bibfield  {author} {\bibinfo {author} {\bibfnamefont {F.}~\bibnamefont
  {Uhlig}},\ }\bibfield  {title} {\bibinfo {title} {Simultaneous block
  diagonalization of two real symmetric matrices},\ }\href
  {https://doi.org/https://doi.org/10.1016/S0024-3795(73)80001-1} {\bibfield
  {journal} {\bibinfo  {journal} {Linear Algebra and its Applications}\
  }\textbf {\bibinfo {volume} {7}},\ \bibinfo {pages} {281} (\bibinfo {year}
  {1973})}\BibitemShut {NoStop}%
\bibitem [{\citenamefont {Maehara}\ and\ \citenamefont
  {Murota}(2010{\natexlab{a}})}]{Maehara2010-1}%
  \BibitemOpen
  \bibfield  {author} {\bibinfo {author} {\bibfnamefont {T.}~\bibnamefont
  {Maehara}}\ and\ \bibinfo {author} {\bibfnamefont {K.}~\bibnamefont
  {Murota}},\ }\bibfield  {title} {\bibinfo {title} {Error-controlling
  algorithm for simultaneous block-diagonalization and its application to
  independent component analysis},\ }\href@noop {} {\bibfield  {journal}
  {\bibinfo  {journal} {JSIAM Letters}\ }\textbf {\bibinfo {volume} {2}},\
  \bibinfo {pages} {131} (\bibinfo {year} {2010}{\natexlab{a}})}\BibitemShut
  {NoStop}%
\bibitem [{\citenamefont {Maehara}\ and\ \citenamefont
  {Murota}(2010{\natexlab{b}})}]{Maehara2010-2}%
  \BibitemOpen
  \bibfield  {author} {\bibinfo {author} {\bibfnamefont {T.}~\bibnamefont
  {Maehara}}\ and\ \bibinfo {author} {\bibfnamefont {K.}~\bibnamefont
  {Murota}},\ }\bibfield  {title} {\bibinfo {title} {A numerical algorithm for
  block-diagonal decomposition of matrix $\ast$-algebras with general
  irreducible components},\ }\href@noop {} {\bibfield  {journal} {\bibinfo
  {journal} {Japan journal of industrial and applied mathematics}\ }\textbf
  {\bibinfo {volume} {27}},\ \bibinfo {pages} {263} (\bibinfo {year}
  {2010}{\natexlab{b}})}\BibitemShut {NoStop}%
\bibitem [{\citenamefont {Murota}\ \emph {et~al.}(2010)\citenamefont {Murota},
  \citenamefont {Kanno}, \citenamefont {Kojima},\ and\ \citenamefont
  {Kojima}}]{Murota2010}%
  \BibitemOpen
  \bibfield  {author} {\bibinfo {author} {\bibfnamefont {K.}~\bibnamefont
  {Murota}}, \bibinfo {author} {\bibfnamefont {Y.}~\bibnamefont {Kanno}},
  \bibinfo {author} {\bibfnamefont {M.}~\bibnamefont {Kojima}},\ and\ \bibinfo
  {author} {\bibfnamefont {S.}~\bibnamefont {Kojima}},\ }\bibfield  {title}
  {\bibinfo {title} {A numerical algorithm for block-diagonal decomposition of
  matrix $\ast$-algebras with application to semidefinite programming},\
  }\href@noop {} {\bibfield  {journal} {\bibinfo  {journal} {Japan Journal of
  Industrial and Applied Mathematics}\ }\textbf {\bibinfo {volume} {27}},\
  \bibinfo {pages} {125} (\bibinfo {year} {2010})}\BibitemShut {NoStop}%
\bibitem [{\citenamefont {Maehara}\ and\ \citenamefont
  {Murota}(2011)}]{Maehara2011}%
  \BibitemOpen
  \bibfield  {author} {\bibinfo {author} {\bibfnamefont {T.}~\bibnamefont
  {Maehara}}\ and\ \bibinfo {author} {\bibfnamefont {K.}~\bibnamefont
  {Murota}},\ }\bibfield  {title} {\bibinfo {title} {Algorithm for
  error-controlled simultaneous block-diagonalization of matrices},\
  }\href@noop {} {\bibfield  {journal} {\bibinfo  {journal} {SIAM Journal on
  Matrix Analysis and Applications}\ }\textbf {\bibinfo {volume} {32}},\
  \bibinfo {pages} {605} (\bibinfo {year} {2011})}\BibitemShut {NoStop}%
\bibitem [{\citenamefont {Irving}\ and\ \citenamefont
  {Sorrentino}(2012)}]{Ir:So}%
  \BibitemOpen
  \bibfield  {author} {\bibinfo {author} {\bibfnamefont {D.}~\bibnamefont
  {Irving}}\ and\ \bibinfo {author} {\bibfnamefont {F.}~\bibnamefont
  {Sorrentino}},\ }\bibfield  {title} {\bibinfo {title} {Synchronization of a
  hypernetwork of coupled dynamical systems},\ }\href@noop {} {\bibfield
  {journal} {\bibinfo  {journal} {Phys. Rev. E}\ }\textbf {\bibinfo {volume}
  {86}},\ \bibinfo {pages} {056102} (\bibinfo {year} {2012})}\BibitemShut
  {NoStop}%
\bibitem [{\citenamefont {Zhang}\ and\ \citenamefont
  {Motter}(2020)}]{Zhang2020}%
  \BibitemOpen
  \bibfield  {author} {\bibinfo {author} {\bibfnamefont {Y.}~\bibnamefont
  {Zhang}}\ and\ \bibinfo {author} {\bibfnamefont {A.~E.}\ \bibnamefont
  {Motter}},\ }\bibfield  {title} {\bibinfo {title} {Symmetry-independent
  stability analysis of synchronization patterns},\ }\href@noop {} {\bibfield
  {journal} {\bibinfo  {journal} {SIAM Review}\ }\textbf {\bibinfo {volume}
  {62}},\ \bibinfo {pages} {817} (\bibinfo {year} {2020})}\BibitemShut
  {NoStop}%
\bibitem [{\citenamefont {Zhang}\ \emph {et~al.}(2021)\citenamefont {Zhang},
  \citenamefont {Latora},\ and\ \citenamefont {Motter}}]{zhang2021unified}%
  \BibitemOpen
  \bibfield  {author} {\bibinfo {author} {\bibfnamefont {Y.}~\bibnamefont
  {Zhang}}, \bibinfo {author} {\bibfnamefont {V.}~\bibnamefont {Latora}},\ and\
  \bibinfo {author} {\bibfnamefont {A.~E.}\ \bibnamefont {Motter}},\ }\bibfield
   {title} {\bibinfo {title} {Unified treatment of synchronization patterns in
  generalized networks with higher-order, multilayer, and temporal
  interactions},\ }\href@noop {} {\bibfield  {journal} {\bibinfo  {journal}
  {Communications Physics}\ }\textbf {\bibinfo {volume} {4}},\ \bibinfo {pages}
  {1} (\bibinfo {year} {2021})}\BibitemShut {NoStop}%
\bibitem [{\citenamefont {Pastor-Satorras}\ and\ \citenamefont
  {Vespignani}(2001)}]{pastor2001epidemic}%
  \BibitemOpen
  \bibfield  {author} {\bibinfo {author} {\bibfnamefont {R.}~\bibnamefont
  {Pastor-Satorras}}\ and\ \bibinfo {author} {\bibfnamefont {A.}~\bibnamefont
  {Vespignani}},\ }\bibfield  {title} {\bibinfo {title} {Epidemic spreading in
  scale-free networks},\ }\href@noop {} {\bibfield  {journal} {\bibinfo
  {journal} {Physical review letters}\ }\textbf {\bibinfo {volume} {86}},\
  \bibinfo {pages} {3200} (\bibinfo {year} {2001})}\BibitemShut {NoStop}%
\bibitem [{\citenamefont {Marder}(2007)}]{marder2007dynamics}%
  \BibitemOpen
  \bibfield  {author} {\bibinfo {author} {\bibfnamefont {M.}~\bibnamefont
  {Marder}},\ }\bibfield  {title} {\bibinfo {title} {Dynamics of epidemics on
  random networks},\ }\href@noop {} {\bibfield  {journal} {\bibinfo  {journal}
  {Physical Review E}\ }\textbf {\bibinfo {volume} {75}},\ \bibinfo {pages}
  {066103} (\bibinfo {year} {2007})}\BibitemShut {NoStop}%
\bibitem [{\citenamefont {Pastor-Satorras}\ \emph {et~al.}(2015)\citenamefont
  {Pastor-Satorras}, \citenamefont {Castellano}, \citenamefont {Van~Mieghem},\
  and\ \citenamefont {Vespignani}}]{pastor2015epidemic}%
  \BibitemOpen
  \bibfield  {author} {\bibinfo {author} {\bibfnamefont {R.}~\bibnamefont
  {Pastor-Satorras}}, \bibinfo {author} {\bibfnamefont {C.}~\bibnamefont
  {Castellano}}, \bibinfo {author} {\bibfnamefont {P.}~\bibnamefont
  {Van~Mieghem}},\ and\ \bibinfo {author} {\bibfnamefont {A.}~\bibnamefont
  {Vespignani}},\ }\bibfield  {title} {\bibinfo {title} {Epidemic processes in
  complex networks},\ }\href@noop {} {\bibfield  {journal} {\bibinfo  {journal}
  {Reviews of modern physics}\ }\textbf {\bibinfo {volume} {87}},\ \bibinfo
  {pages} {925} (\bibinfo {year} {2015})}\BibitemShut {NoStop}%
\bibitem [{\citenamefont {Achlioptas}\ \emph {et~al.}(2009)\citenamefont
  {Achlioptas}, \citenamefont {D'Souza},\ and\ \citenamefont
  {Spencer}}]{achlioptas2009explosive}%
  \BibitemOpen
  \bibfield  {author} {\bibinfo {author} {\bibfnamefont {D.}~\bibnamefont
  {Achlioptas}}, \bibinfo {author} {\bibfnamefont {R.~M.}\ \bibnamefont
  {D'Souza}},\ and\ \bibinfo {author} {\bibfnamefont {J.}~\bibnamefont
  {Spencer}},\ }\bibfield  {title} {\bibinfo {title} {Explosive percolation in
  random networks},\ }\href@noop {} {\bibfield  {journal} {\bibinfo  {journal}
  {science}\ }\textbf {\bibinfo {volume} {323}},\ \bibinfo {pages} {1453}
  (\bibinfo {year} {2009})}\BibitemShut {NoStop}%
\bibitem [{\citenamefont {Friedman}\ and\ \citenamefont
  {Landsberg}(2009)}]{friedman2009construction}%
  \BibitemOpen
  \bibfield  {author} {\bibinfo {author} {\bibfnamefont {E.~J.}\ \bibnamefont
  {Friedman}}\ and\ \bibinfo {author} {\bibfnamefont {A.~S.}\ \bibnamefont
  {Landsberg}},\ }\bibfield  {title} {\bibinfo {title} {Construction and
  analysis of random networks with explosive percolation},\ }\href@noop {}
  {\bibfield  {journal} {\bibinfo  {journal} {Physical review letters}\
  }\textbf {\bibinfo {volume} {103}},\ \bibinfo {pages} {255701} (\bibinfo
  {year} {2009})}\BibitemShut {NoStop}%
\bibitem [{\citenamefont {Guillaume}\ \emph {et~al.}(2004)\citenamefont
  {Guillaume}, \citenamefont {Latapy},\ and\ \citenamefont
  {Magnien}}]{guillaume2004comparison}%
  \BibitemOpen
  \bibfield  {author} {\bibinfo {author} {\bibfnamefont {J.-L.}\ \bibnamefont
  {Guillaume}}, \bibinfo {author} {\bibfnamefont {M.}~\bibnamefont {Latapy}},\
  and\ \bibinfo {author} {\bibfnamefont {C.}~\bibnamefont {Magnien}},\
  }\bibfield  {title} {\bibinfo {title} {Comparison of failures and attacks on
  random and scale-free networks},\ }in\ \href@noop {} {\emph {\bibinfo
  {booktitle} {International Conference on Principles of Distributed
  Systems}}}\ (\bibinfo {organization} {Springer},\ \bibinfo {year} {2004})\
  pp.\ \bibinfo {pages} {186--196}\BibitemShut {NoStop}%
\bibitem [{\citenamefont {Liu}\ \emph {et~al.}(2012)\citenamefont {Liu},
  \citenamefont {Wang}, \citenamefont {Lai},\ and\ \citenamefont
  {Wang}}]{liu2012cascading}%
  \BibitemOpen
  \bibfield  {author} {\bibinfo {author} {\bibfnamefont {R.-R.}\ \bibnamefont
  {Liu}}, \bibinfo {author} {\bibfnamefont {W.-X.}\ \bibnamefont {Wang}},
  \bibinfo {author} {\bibfnamefont {Y.-C.}\ \bibnamefont {Lai}},\ and\ \bibinfo
  {author} {\bibfnamefont {B.-H.}\ \bibnamefont {Wang}},\ }\bibfield  {title}
  {\bibinfo {title} {Cascading dynamics on random networks: Crossover in phase
  transition},\ }\href@noop {} {\bibfield  {journal} {\bibinfo  {journal}
  {Physical Review E}\ }\textbf {\bibinfo {volume} {85}},\ \bibinfo {pages}
  {026110} (\bibinfo {year} {2012})}\BibitemShut {NoStop}%
\bibitem [{\citenamefont {Devlin}\ and\ \citenamefont
  {Treloar}(2009)}]{devlin2009evolution}%
  \BibitemOpen
  \bibfield  {author} {\bibinfo {author} {\bibfnamefont {S.}~\bibnamefont
  {Devlin}}\ and\ \bibinfo {author} {\bibfnamefont {T.}~\bibnamefont
  {Treloar}},\ }\bibfield  {title} {\bibinfo {title} {Evolution of cooperation
  through the heterogeneity of random networks},\ }\href@noop {} {\bibfield
  {journal} {\bibinfo  {journal} {Physical Review E}\ }\textbf {\bibinfo
  {volume} {79}},\ \bibinfo {pages} {016107} (\bibinfo {year}
  {2009})}\BibitemShut {NoStop}%
\bibitem [{\citenamefont {Restrepo}\ \emph {et~al.}(2006)\citenamefont
  {Restrepo}, \citenamefont {Ott},\ and\ \citenamefont
  {Hunt}}]{restrepo2006synchronization}%
  \BibitemOpen
  \bibfield  {author} {\bibinfo {author} {\bibfnamefont {J.~G.}\ \bibnamefont
  {Restrepo}}, \bibinfo {author} {\bibfnamefont {E.}~\bibnamefont {Ott}},\ and\
  \bibinfo {author} {\bibfnamefont {B.~R.}\ \bibnamefont {Hunt}},\ }\bibfield
  {title} {\bibinfo {title} {Synchronization in large directed networks of
  coupled phase oscillators},\ }\href@noop {} {\bibfield  {journal} {\bibinfo
  {journal} {Chaos: An Interdisciplinary Journal of Nonlinear Science}\
  }\textbf {\bibinfo {volume} {16}},\ \bibinfo {pages} {015107} (\bibinfo
  {year} {2006})}\BibitemShut {NoStop}%
\bibitem [{\citenamefont {Liu}\ \emph {et~al.}(2011)\citenamefont {Liu},
  \citenamefont {Slotine},\ and\ \citenamefont
  {Barab{\'a}si}}]{liu2011controllability}%
  \BibitemOpen
  \bibfield  {author} {\bibinfo {author} {\bibfnamefont {Y.-Y.}\ \bibnamefont
  {Liu}}, \bibinfo {author} {\bibfnamefont {J.-J.}\ \bibnamefont {Slotine}},\
  and\ \bibinfo {author} {\bibfnamefont {A.-L.}\ \bibnamefont {Barab{\'a}si}},\
  }\bibfield  {title} {\bibinfo {title} {Controllability of complex networks},\
  }\href@noop {} {\bibfield  {journal} {\bibinfo  {journal} {Nature}\ }\textbf
  {\bibinfo {volume} {473}},\ \bibinfo {pages} {167} (\bibinfo {year}
  {2011})}\BibitemShut {NoStop}%
\end{thebibliography}
\end{document}